\documentclass[11pt, twoside, a4paper]{article}
\usepackage{a4wide}

\usepackage[latin1]{inputenc}
\usepackage{amssymb}
\usepackage{amsfonts}
\usepackage{amsthm}
\usepackage{latexsym}
\usepackage{epsfig}
\usepackage{float}
\usepackage{graphicx}
\usepackage{array}
\usepackage{multirow}
\usepackage{alltt}
\usepackage{moreverb}

\newcommand{\Cond}[2]{#1_{\rm #2}}

\newcommand{\Var}[1]{\mathbf{var}(#1)}
\newcommand{\Red}[2]{\mathrm{red}_{#1}^{#2}}
\newcommand{\restrict}[2]{#1|_{#2}}
\newlength{\indentcode}
\newcommand{\Code}[1]{{\tt #1}}

\newcommand{\B}[1]{{\bfseries #1}}

\renewcommand{\L}[1]{{\sffamily #1}}

\def\oadymppac{OADymPPaC}
\def\discipl{DiSCiPl}
\def\tell{\L{tell}}

\def\told{\L{told}}
\def\reduce{\L{reduce}}
\def\suspend{\L{suspend}}
\def\true{\L{true}}
\def\reject{\L{reject}}
\def\wakeup{\L{wake-up}}
\def\select{\L{select}}

\def\clpfd{\L{clp(fd)}}

\def\Cdiff{\#\#}
\def\Cssup{\#>}

\newenvironment{myitemize}
        {\begin{list}
                {$\bullet$}{                    
                \leftmargin 1cm
                \rightmargin 0cm                
                \topsep 0cm                     
                \itemsep 0cm                    
                \parsep 0cm                     
                }  
        }{\end{list}}

\newtheorem{definition}{Definition}[section]

\begin{document}
\pagestyle{myheadings}
\markboth{WLPE'01}{Prototyping \clpfd{} Tracers}


\title{Prototyping \clpfd{} Tracers: a Trace Model \\and an
  Experimental Validation Environment \thanks{This work is partly
    supported by \oadymppac{}, a French RNTL project~\cite{oadimpac}.}
\footnote{In A. Kusalik (ed), Proceedings of the Eleventh Workshop
    on Logic Programming Environments (WLPE'01), December 1, 2001,
    Paphos, Cyprus.  Computer Research Repository
    (http://www.acm.org/corr/), cs.PL/0111043; 
       whole proceedings: cs.PL/0111042.}}
\author{Ludovic Langevine, Pierre Deransart, Mireille Ducass\'{e} and Erwan Jahier\\
  INRIA/INSA, INRIA, IRISA/INSA, IRISA/IFSIC\\
  \\{\small\{Ludovic.Langevine Pierre.Deransart\}@inria.fr,
    \{Mireille.Ducasse Erwan.Jahier\}@irisa.fr}}
\date{}

\maketitle

\begin{abstract}
Developing and maintaining CLP programs requires visualization and
explanation tools. However, existing tools are built in an ad hoc
way. Therefore porting tools from one platform to another is very
difficult. We have shown in previous work that, from a fine-grained
execution trace, a number of interesting views about logic program
executions could be generated by trace analysis.

In this article, we propose a trace model for constraint solving by
narrowing. This trace model is the first one proposed for \clpfd{} and
does not pretend to be the ultimate one.  We also propose an
instrumented meta-interpreter in order to experiment with the
model. Furthermore, we show that the proposed trace model contains the
necessary information to build known and useful execution views. This
work sets the basis for generic execution analysis of \clpfd{} programs.

\cite{oad121} is a comprehensive version of this paper.
\end{abstract}

\section{Introduction}

Developing and maintaining CLP programs benefits from visualization and
explanation tools such as the ones designed by the \discipl{} European
project \cite{deransart00}. For example, the CHIP search-tree tool
\cite{simonis00} helps users understand the effect of a search
procedure on the search space; and the S-Box model \cite{goualard00}
allows users to inspect the store with graphical and hierarchical
representations.

However, existing tools are built in an ad hoc way. Therefore porting
tools from one platform to another is very difficult. One has to
duplicate the whole design and implementation effort every time.

We have shown in previous work that, from a fine-grained execution
trace, a number of interesting views about logic program executions
could be generated by trace analysis~\cite{ducasse99b,jahier00c}.
Here we want to generate a generic trace that can be used by several
debugging tools. Each tool gives a useful view of the \clpfd{}
execution (e.g. the labelling tree or the domain state evolution
during a propagation stage). Each tool selects the relevant data in
the generic trace in order to build its execution view.  Therefore,
the trace is supposed to contain all potentially useful information.
The tools can be developed simultaneously and reused on different
platforms\,\footnote{ Our \oadymppac{} partners will use our trace
  model to build tools in the CHIP~\cite{chip98},
  CLAIRE~\cite{caseau99} and GNU Prolog~\cite{gnuprolog} platforms.}.
This approach assumes that there is a generic trace model. In order to
reach this objective, we start with a general model of finite domain
constraint solving which allows to define generic trace events.

So far, no fine-grained tracers exist for constraint solvers.
Implementing a formally defined tracer is a delicate task, if even at
all possible. This is especially true in the context of constraint
solving where the solvers are highly optimized. Therefore, before
going for a ``real'' implementation, it is essential to elaborate
trace models and experiment them.

In order to rigorously define the constraint solving trace model, and
as we did for Prolog \cite{jahier00}, we use an operational semantics
based on \cite{benhamou94} and \cite{ferrand00}. We define execution
events of interest with respect to this semantics. On one hand, this
formal approach prevents the fined-grained trace model to be platform
dependent.  On the other hand, the consistence between the model and
concrete \clpfd{} platforms has to be validated.

In order to develop a first experimentation, the formal model is
implemented as an instrumented meta-interpreter\,\footnote{The
  meta-interpreter approach to trace \clpfd{} program executions has
  already been used in the APT tool by Carro and
  Hermenegildo~\cite{carro00}. However, APT does not access the
  propagation details.  We propose a more informative trace.} which
exactly reflects it.

The proposed instrumented meta-interpreter is useful to
experimentally validate the model. It is based on well known Prolog
meta-interpretation techniques for the Prolog part and
on the described operational semantic for the constraint part.

The ultimate goal of the meta-interpreter (as in \cite{jahier00}) is
to provide an executable specification of traces. The traces generated
by the meta-interpreter could then be used to validate a real (and
efficient) tracer.  At this stage, efficiency of the meta-interpreter
is not a key issue; it is used as a prototype rather than as an
effective implementation.

Such a tracer may generate a large volume of data. The generic trace
must be analyzed and filtered in order to condense this data down to
provide the sufficient information to the debugging tools.  To push
further the validation we have mimicked a trace analyzer \emph{à la}
Opium~\cite{ducasse99} which allow to obtain different kinds of views
of the execution.  This is a way to show that the proposed trace model
contains the necessary information to reproduce several existing
\clpfd{} debugging tools.

In this article, due to lack of space we concentrate on the constraint
solving part, a trace of the logic programming part ``\emph{à la} Byrd''
\cite{byrd80} could be integrated in the meta-interpreter (see for
example \cite{ducasse93c}).

The contribution of this article is first to rigorously define a
trace model for constraint solving and to provide an environment to
experimentally validate it. This sets the basis for generic execution
analysis of \clpfd{} programs.

In the following, Section~\ref{execution:model:section} formally
defines an execution model of \clpfd{} narrowing. In particular, it
defines a 8 steps operational semantics of constraint solving. These
steps are the basis for the trace format defined in
Section~\ref{trace:section}. Section~\ref{meta:interpreter:section}
explains how to build an instrumented meta-interpreter in order to
build an experimental environment. Section~\ref{validation:section}
briefly describes experiments with the proposed trace and our trace
analyzer. Finally, Section~\ref{conclusion:section} discusses the
content of the trace with respect to existing debugging tools.


\section{Operational Semantics of Constraint Programming}
\label{execution:model:section}

In this section, we propose an execution model of (finite domain)
constraint programming which is language independent.  The operational
semantics of constraint programming results from the combination of
two paradigms: control and propagation. The control part depends on
the programming language in which the solver is embedded, and the
propagation corresponds to narrowing. Although the notions introduced
here are essentially language independent, we will illustrate them in
the context of \clpfd{}.

\subsection{Basic Notations}

In the rest of the paper, ${\cal P}(A)$ denotes the power set of $A$;
$\restrict{r}{w}$ denotes the \emph{restriction} of the relation $r
\subseteq A \times B$ to $w \subseteq A$:  
$\;\;\;\restrict{r}{w} = \{(x, y) \mid x \in w \wedge (x, y) \in r \}$.

The following notations are attached to variables and constraints:
$\cal V$ is the set of all the finite domain variables of the problem;
$\cal D$ is a finite set containing all possible values for
    variables in $\cal V$;
$D$ is a function $D: \cal V \rightarrow \cal P(D)$, which
    associates to each variable $x$ its current domain, denoted by
    $D_x$;
$min_x$ and $max_x$ are respectively the lower and upper bounds of
    $D_x$;
$\cal C$ is the set of the problem constraints;
 $\mathbf{var}$ is a function $\cal C \rightarrow P(V)$, which
    associates to each constraint $C \in {\cal C}$ the set of
    variables of the constraint. 

\subsection{Reduction Operators}
\label{reduction:operator:section}

Many explanation tools focus on domain reduction (see for
example~\cite{ferrand00,jussien00}). When constraints are propagated,
the evolution of variable domains is a sequence of withdrawals of
inconsistent values. At each step, for a given constraint, a set of
inconsistent values is withdrawn from one (and only one) domain. These
values can be determined by algorithms such as ``node consistency'',
``arc consistency'', ``hyper-arc consistency'' or ``bounds
consistency'' described for example by Marriott and
Stuckey~\cite{marriott98}.

Following Ferrand~et~al~\cite{ferrand00}, we define reduction
operators. The application of all the reduction operators of a
constraint gives the \emph{narrowing operator} introduced by
Benhamou~\cite{benhamou94}.

\begin{definition}[Reduction operator]
  A \emph{reduction operator} $red^x_C$ is a function attached to a
  constraint $C$ and a variable $x$. Given the domains of all the
  variables used in $C$, it returns the domain $D_x$ without the
  values of $x$ which are inconsistent with the domains of the other
  variables. The set of withdrawn values is denoted $W_x$.
  
$red_C^x(\restrict{D}{\Var{C}}) = D_x - W_x$.\qed
\end{definition}

There are as many reduction operators attached to $C$ as variables in
$\Var{C}$.  In general, for efficiency reasons, a reduction operator
does not withdraw all inconsistent values.

A simple example of reduction operator for $C \equiv x > n$, where $n$
is a given integer, is $red^x_C(\restrict{D}{\{x\}}) = D_x - \{v \mid
v \in D_x \wedge v \leq n\}$.

The evolution of the domains can be viewed as a sequence of applications of
reduction operators attached to the constraints of the store.  Each
operator can be applied several times until the computation reaches a
fix-point~\cite{ferrand00}. This fix-point is the set of final domain states.

\begin{figure}[t]
  \begin{center}
    \includegraphics[width=14.5cm]{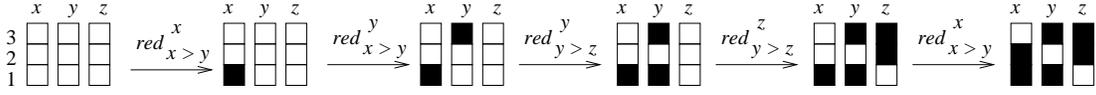}
    \caption{Application of reductions to the system $\{x> y; y >z\}$.}
    \label{FIG-REDUCTION}
  \end{center}
\end{figure}

An example of computation with reduction operators is shown in
Figure~\ref{FIG-REDUCTION}. There are three variables $x$, $y$ and $z$
and two constraints, $x > y$ and $y > z$.
At the beginning, $D_x = D_y = D_z = \{1, 2, 3\}$, represented by
three columns of white squares. Considering the first constraint, it
appears that $x$ cannot take the value ``1'', because otherwise there
would be no value for $y$ such that $x > y$; $\Red{x \mathrm{>}
y}{x}$ withdraws this inconsistent value from $D_x$. This withdrawal
is marked with a black square. In the same way, $\Red{x \mathrm{>}
y}{y}$ withdraws the value 3 from the domain of $y$. Then, considering
the constraint $y>z$, the operators $\Red{y
\mathrm{>} z}{y}$ and $\Red{y \mathrm{>} z}{z}$ withdraw
respectively the sets $\{1\}$ and $\{2, 3\}$ from $D_y$ and
$D_z$. Finally, a second application of $\Red{x \mathrm{>} y}{x}$
reduces $D_x$ to the singleton $\{3\}$.  The fix-point is reached. The
final solution is: $\{x = 3, y = 2, z = 1\}$.

\subsection{Awakening and Solved Conditions}
\label{conditions:section}

An essential notion of constraint propagation is the \emph{propagation
  queue}. This queue contains all the constraints whose reduction
operators have to be applied. At each step of the propagation, a
constraint is selected in the propagation queue, according to a given
strategy depending on implementation (for example a constraint with
more variables first). The reduction operators of the selected
constraint are applied. These applications can make new domain
reductions. When a variable domain is updated, the system puts in the
queue all the constraints where this variable appears. When the queue
becomes empty, a fix-point is reached and the propagation ends. Thus,
there are three fundamental operations: \emph{selection} from the
queue, \emph{reduction} of variable domains and \emph{awakening} of
constraints.

It is actually not necessary to wake a constraint on each update of
its variable domains. For example, it is irrelevant to wake the
constraint $x > y$ at each modification of $D_x$ or
$D_y$. The reduction operators of this constraint are unable to
withdraw a value in any domain if neither the upper bound of $D_x$
nor the lower bound of $D_y$ is updated. Hence, it is sufficient to
wake the constraint only when one of those particular modifications
occurs.

\begin{definition}[Awakening condition]
  The \emph{awakening condition} of a constraint $C$ is a predicate
  depending upon the modifications of $\restrict{D}{\Var{C}}$. This
  condition holds when a new value withdrawal can be made by the
  constraint reduction operators.  The condition is optimal when it
  holds \emph{only when} a new value withdrawal can be made.
  
  The awakening condition of $C$ is denoted by
  $\mathrm{awake\_cond}(C)$.\qed
\end{definition}
The actual awakening conditions are often a compromise between the
cost of their computation and how many awakenings they prevent.
  
\paragraph{}
Another condition type permits irrelevant awakenings to be
prevented. Let us consider a constraint $C$; if the domains of its
variables are such that no future value withdrawals can invalidate the
constraint satisfaction, the constraint is said to be
\emph{solved}. In that case, the reduction operators of this
constraint cannot make any new value withdrawal anymore, unless the
system backtracks to a former point.  For example, if the two domains
$D_x$ and $D_y$ are such that $D_x \cap D_y = \emptyset$, it is
useless to apply the reduction operators of $x \neq y$. Thus, it is
useless to wake a solved constraint.

\begin{definition}[Solved condition]
  The \emph{solved condition} of a constraint $C$ is a predicate
  depending upon the state of $\restrict{D}{\Var{C}}$. This condition
  holds only when the domains are such that $C$ is solved. 

The solved condition of $C$ is denoted by $\mathrm{solved\_cond}(C)$.\qed
\end{definition}
We can note that a sufficient condition is that all variables in
$\Var{C}$ are ground to some values satisfying the constraint.

In the rest of the paper, a primitive constraint is defined by those
three characteristics: its reduction operators, awakening
condition and solved condition.

\subsection{Structure of the Constraint Store}
\label{store:decomposition:section}

The \emph{constraint store} $\cal S$ is the set of all constraints
taken into account by the computation at a given moment. When the
computation begins, the store is empty. Then, constraints are
individually added or withdrawn according to the control part. In the
store, constraints may have different status and the store can be
partitioned into five subsets denoted $A, S, Q, T, R$.  

\begin{myitemize}
\item $A$ is the set of \emph{active} constraints. It is either a
  singleton\footnote{This restriction could  be alleviated to handle multiple active
    constraints, for example to handle unification viewed as equality
    constraints on Herbrand's domains.}  (only one constraint is
  active) or empty (no constraint is active). The reduction operators
  associated to the active constraint will perform the reductions of
  variable domains.
\item $S$ is the set constraints which are said
\emph{suspended}, namely they are waiting to be ``woken'' and put into
$Q$ in the case of domain modifications of some of their variable
(none has empty domain).
\item $Q$ is the set of constraints in the \emph{propagation
    queue}\,\footnote{The term \emph{queue} comes from standard usage. It is in fact a set.}, it contains the constraints waiting to be activated. In
  order to reach the fix-point all reduction operators associated to
  these constraints must be considered.
\item $T$ is the set of solved constraints (\emph{true}), namely the
constraints which hold whatever future withdrawals will be made.
\item $R$ is the set of rejected constraints, i.e. the constraints
for which the domain of at least one variable is empty. In practice it
is empty or a singleton, since as soon as there is one constraint in
$R$ the store is considered as ``unsatisfiable'' and the computations
will continue according to the control.
\end{myitemize}

\subsection{Propagation}
\label{propagation:section}

\begin{figure}
  \begin{center}
    \includegraphics[width=12cm]{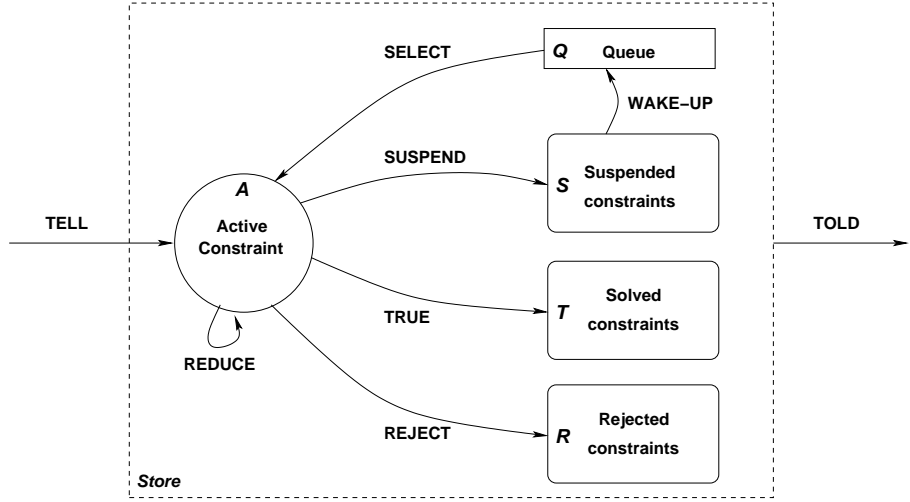}
  \end{center}
  \caption{Events related to the structured store}
\label{FIG-STORE}
\end{figure}

The evolution of the store can now be described as state transition
functions or ``events'' acting in the store, in the style of
Guerevitch's evolving algebras~\cite{guerevitch91}. This is
illustrated in Figure~\ref{FIG-STORE}.  When propagation begins, there
is an active constraint. The active constraint applies its reduction
operators as far as possible. Each application of a reduction operator
is a \reduce{} event. It narrows the domain of one and only one
variable.  If a domain becomes empty, the constraint is rejected
(\reject{} event). A rejected constraint is a sign of failure. If no
failure occurs and no other reduction can be made, the constraint is
either solved (\true{}) or suspended (\suspend{}). A solved constraint
will not be woken anymore.  A suspended constraint will be woken as
soon as its awakening condition holds. On awakening, the constraint is
put in the propagation queue (\wakeup{} event). When there is no
active constraint, one is selected in the queue (\select{}) and
becomes active. If the queue is empty, the propagation ends.

\newcommand{\regle}[4]{\makebox[2.0cm][r]{\B{#1}}\quad\quad$\displaystyle\frac{#2}{#3}\quad #4$}%
\def\saut{\\[.2cm]}%
\def\et{~\wedge~}
\def\virg{,~}
\begin{figure}
\small
      \regle{select}
            {\exists C \in Q \et A = \emptyset \et R = \emptyset}
            {Q \gets Q - \{C\}\virg A \gets \{C\}}
            {}\\
            \saut
      \regle{reject}
            {\exists C \in A, \exists x \in \Var{C} \cdot (D_x = \emptyset)}
            {A \gets A - \{C\}\virg R \gets \{C\}}
            {}\\
            \saut
      \regle{wake-up}
            {\exists C \in S \cdot \mathrm{awake\_cond}(C) \et R = \emptyset}
            {S \gets S - \{C\}\virg Q \gets Q \cup \{C\}}
            {}\\
            \saut
      \regle{reduce}
            {\exists C \in A, \exists x \in \Var{C} \cdot (W_x \neq \emptyset) \et R = \emptyset}
            {D_x \gets D_x - W_x}
            {W_x = D_x - red^x_C(\restrict{D}{\Var{C}})}\\
            \saut
      \regle{true}
            {\exists C \in A \cdot \mathrm{solved\_cond}(C) \et R = \emptyset}
            {A \gets A - \{C\}\virg T \gets T \cup \{C\}}
            {}\\
            \saut
      \regle{suspend}
            {\exists C \in A \et R = \emptyset}
            {A \gets A - \{C\}\virg S \gets S \cup \{C\}}
            {}
            \saut
  
    \caption{Propagation events. ~~~Rule format : {\small \B{Name}\quad$\displaystyle\frac{Conditions}{Actions}~Definitions$}}
        \label{FIG-RULES}

\vspace{0.5cm}
      \regle{tell(C)}
            {A = \emptyset}
            {push~\langle\{C\}, S, Q, T, R, D\rangle}
            {}\\
            \saut
      \regle{told}
            {}
            {\quad\quad pop\quad\quad}
            {}
    \caption{Control events}
    \label{FIG-RULES2}
\end{figure}

The propagation is completely defined by the rules given in
Figure~\ref{FIG-RULES}.  The rules are applied in the order in which
they are given (from top to bottom).  Each rule specifies an event
type.  An event modifies the system state: $\langle
A,S,Q,T,R,D\rangle$.  An event occurs when its pre-conditions hold and
no higher-priority event is possible.  The rule priority prevents
redundant conditions.  For example, a \suspend{} event is made only
when no \true{} event is possible.  In the same way, if a \reduce{}
event wakes some constraints, all \wakeup{} events are performed
before any other \reduce{}.  The rule system still contains some
indeterminism: the choice of $C$ in the rules \select{} and \wakeup{},
and the choice of $x$ in the rule \reduce{} depends on the solver
strategy.

The rules are applied until all the constraints are in $T$, $R$ and
$S$.  No further application is possible.
If $R$ is non empty, the store is ``rejected''.  Otherwise, it may be
considered that a (set of) solution(s) has been obtained.  All
constraints in $T$ are already solved. Therefore, any tuple of values
of $\cal V$ which satisfies the constraints in $S$ is a solution.  The
way the computation will continue depends of the control.

\subsection{Control}
\label{control:section}

The evolution of the store described so far assumes that the set of
constraints in the store is invariant, i.e., only their status is
modified.
The host constraint programming language provides the way to build the
store and to manage it, with possible interleaving of constraint
management and propagation steps.

In order to remain as independent as possible from the host language,
we restrict the control part to two events: the first one adds a
constraint $C$ into the store (\tell{}$(C)$ event) and the second one
restores the store and the domains in some previous state (\told{} event).

These events act on the store as illustrated in
Figure~\ref{FIG-STORE}.
The \tell{} event puts a new constraint in the store as the active one.

In \clpfd{} the control part is particularly simple and can be
described by a (control) stack of states of the system~\footnote{This
  corresponds to the description of the visit of a standard
  search-tree of Prolog~\cite{alipie96}.}.  The state contains the
current store and the domains. The propagation phase is always
performed in the state on top of the control stack.  Each \tell{}
event includes a constraint into the store, push the new state on the
stack and is followed by a complete propagation phase.  When the
fix-point is reached either a new \tell{($C'$)} is performed, or a
\told{}. In the later case the stack is popped, thus the former state
is restored.  As long as the control stack is not empty, new \tell{}
events can be performed on choice points and correspond to alternative
computations.

This semantics is formalized by the two rules of Figure~\ref{FIG-RULES2}.
The $push$ and $pop$ operations work on the control stack.
With a different host language, the same \told{} event could be used but
with a possibly different meaning.

\section{Trace Definition}
\label{trace:section}

An execution is represented by a trace which is a sequence of events.
An event corresponds to an elementary step of the execution. It is a
tuple of attributes.  Following the notation of Prolog traces, the
types of events are called \emph{ports}. 

Most attributes are common to all events. For some ports, specific
attributes are added. For example, \reduce{} events have two
additional attributes: the withdrawn values $W_x$, for the variable
$x$ whose domain is reduced, and the types of updates made, such a
modification of the domain upper bound.  The attributes are as
follows.

\begin{quotation}
\noindent
\B{Attributes for all events}
\begin{myitemize}
\item {\tt chrono}: the event number (starting with 1);

\item {\tt depth}: the depth of the execution, starting with 0, it is
incremented at each \tell{} and decremented at each  \told{};

\item {\tt port}: the event type as presented in
Sections~\ref{propagation:section} and \ref{control:section}: one of
\reduce, \wakeup, \suspend, \true, \reject, \select{}, \tell{} and
\told;

\item {\tt constraint}: the concerned constraint, represented by a quadruple:
        \begin{myitemize}
        \item a unique identifier generated at its \tell{};
        \item an abstract representation, identical to the source formulation of the program;
        \item an concrete representation (e.g. \Code{diffN(X, Y, N)} for
        \Code{X \Cdiff{} Y + N} or \Code{X~-~N~\Cdiff{}~Y}) ;
        \item the invocation context, namely the
Prolog goal from which  the \tell{} is performed.
        \end{myitemize}

\item {\tt domains}: the value of the variable domains before the event occurs;

\item {\tt store}: the content of the constraint store represented by
the 5 components described in
section~\ref{store:decomposition:section}. Each set of constraints is
represented by a list of pairs (constraint identifier, external
representation):
        \begin{myitemize}
        \item {\tt store\_A} the set of \emph{active} constraints;
        \item {\tt store\_S} the set of \emph{suspended} constraints;
        \item {\tt store\_Q} the \emph{propagation queue};
        \item {\tt store\_T} the set of \emph{solved} constraints;
        \item {\tt store\_R} the set of \emph{rejected} constraints.
        \end{myitemize}
\end{myitemize}
\noindent
\B{Specific attributes for \reduce{} events} 
\begin{myitemize}
\item  {\tt withdrawn}: The withdrawn domain
\item {\tt update}: The list of updates, an update is of the form
\Code{(variable -> type)}, where \Code{type} can be one of
\Code{ground, any, min, max}, see Section~\ref{meta:interpreter:section}
for further explanation;
\end{myitemize}
\noindent    
\B{Specific attribute for \wakeup{} events} 
\begin{myitemize}
\item  {\tt cause}: The verified part of the awakening condition
\end{myitemize}

\end{quotation}

\begin{figure}\small
\Code{sorted([X, Y, Z]):-}\\
\makebox[.5cm]{}\Code{[X, Y, Z] :: 1..3,}~~~~~~~~~~\emph{\% At the beginning,  $D_x = D_y = D_z = [1..3]$}\\
\makebox[.5cm]{}\Code{X \Cdiff{} Y, X \Cssup{}= Y, Y \Cssup{}
Z,}~~\emph{\% 3 constraints~: $x \neq y$, $x \geq y$ and $y > z$}\\
\makebox[.5cm]{}\Code{labelling([X, Y, Z]).}~~~~~~~\emph{\% labelling phase, with a ``first fail'' strategy}\\

\begin{minipage}[t]{.45\linewidth}\footnotesize
\begin{alltt}
 1 [1] Tell    X##Y  X:[1,2,3] Y:[1,2,3]
 2 [1] Suspend X##Y  X:[1,2,3] Y:[1,2,3]
 3 [2] Tell    X#>=Y X:[1,2,3] Y:[1,2,3]
 4 [2] Suspend X#>=Y X:[1,2,3] Y:[1,2,3]
 5 [3] Tell    Y#>Z  Y:[1,2,3] Z:[1,2,3]
 6 [3] Reduce  Y#>Z  Y:[1,2,3] Z:[1,2,3] Y[1]
 7 [3] Wake-up X#>=Y X:[1,2,3] Y:[2,3]
 8 [3] Reduce  Y#>Z  Y:[2,3]   Z:[1,2,3] Z[3]
 9 [3] Suspend Y#>Z  Y:[2,3]   Z:[1,2]
10 [3] Select  X#>=Y X:[1,2,3] Y:[2,3]
11 [3] Reduce  X#>=Y X:[1,2,3] Y:[2,3]   X[1]
12 [3] Suspend X#>=Y X:[2,3]   Y:[2,3]
13 [4] Tell    X#=2  X:[2,3]
14 [4] Reduce  X#=2  X:[2,3]             X[3]
15 [4] Wake-up X#>=Y X:[2]     Y:[2,3]
16 [4] Wake-up X##Y  X:[2]     Y:[2,3]
17 [4] True    X#=2  X:[2]
18 [4] Select  X#>=Y X:[2]     Y:[2,3]
19 [4] Reduce  X#>=Y X:[2]     Y:[2,3]   Y[3]
20 [4] Wake-up Y#>Z  Y:[2]     Z:[1,2]
\end{alltt}  
\end{minipage}%
\hfill
\begin{minipage}[t]{.45\linewidth}\footnotesize
\begin{alltt}
21 [4] True    X#>=Y  X:[2]     Y:[2]
22 [4] Select  X##Y   X:[2]     Y:[2]
23 [4] Reduce  X##Y   X:[2]     Y:[2]   X[2]
24 [4] Reject  X##Y   X:[]      Y:[2]
25 [4] Told    X#=2   X:[]
26 [4] Tell    X#=3   X:[2,3]
27 [4] Reduce  X#=3   X:[2,3]           X[2]
28 [4] Wake-up X##Y   X:[3]     Y:[2,3]
29 [4] True    X#=3   X:[3]
30 [4] Select  X##Y   X:[3]     Y:[2,3]
31 [4] Reduce  X##Y   X:[3]     Y:[2,3] Y[3]
32 [4] Wake-up Y#>Z   Y:[2]     Z:[1,2]
33 [4] True    X##Y   X:[3]     Y:[2]
34 [4] Select  Y#>Z   Y:[2]     Z:[1,2]
35 [4] Reduce  Y#>Z   Y:[2]     Z:[1,2] Z[2]
36 [4] True    Y#>Z   Y:[2]     Z:[1]
37 [4] Told    X#=3   X:[3]
38 [3] Told    Y#>Z   Y:[2,3]   Z:[1,2]
39 [2] Told    X#>=Y  X:[1,2,3] Y:[1,2,3]
40 [1] Told    X##Y   X:[1,2,3] Y:[1,2,3]
\end{alltt}
\end{minipage}
{\footnotesize\begin{alltt}

chrono     = 14
depth      =  4
port       = REDUCE
constraint = (4, X#=2, assign(var(1, X), 2), labelling([X, Y, Z]))
domains    = [X::[2..3], Y::[2..3], Z::[1..2]]
withdrawn  = X::[3]
update     = [X->any, X->ground, X->max]
store_A    = [(4, X#=2)]                    store_S = [(2, X#>=Y), (3, Y#>Z), (1, X##Y)]
store_Q    = []             store_T = []    store_R = []   

chrono     = 16 
depth      =  4
port       = WAKE-UP
constraint = (1, X##Y, diff(var(1, X), var(2, Y)), sorted([X, Y, Z]))
domains    = [X::[2], Y::[2..3], Z::[1..2]]
cause      = [X->ground]
store_A    = [(4, X#=2)]                    store_S    = [(3, Y#>Z), (1, X##Y)]  
store_Q    = [(2, X#>=Y)]   store_T = []    store_R = []
\end{alltt}}

\caption{A trace of the execution of program \Code{sorted([X, Y,
Z])}, all events are present with attributes 
(event number, [depth], constraint $C$,
$\restrict{D}{\Var{C}}$). Events \#14 and \#16 are displayed with 
all their attributes.}
\label{trace:example:figure}

\end{figure}

The attributes are numerous and contain large chunks of
information. Indeed, they aim at providing useful information to
automatic trace analysis programs. The more contents the better. In a
default display for users, only some attributes would be
chosen. Furthermore, and as in Opium~\cite{ducasse99} the trace
analysis will be mainly done on the fly, only the attributes relevant
to a given analysis will be retrieved, and {\bf no} trace will be
stored. Therefore there is no a priori restriction on the number and
size of attributes.

Figure~\ref{trace:example:figure} shows the source code of program
\Code{sorted(L)}. The program sorts three numbers between 1 and 3 in a
very naive way. Following the convention of many systems, constraints
operators are prefixed by a ``\Code{\#}''.  The figure also shows a
trace of the execution. All events are listed but only with a few
event attributes: the event number and port, the constraint concerned
by the event and its variable domains. At \reduce{} events, the
variable whose domain is being reduced as well as the withdrawn values
are added.

The first two constraints are entered (\tell{}) and suspended without
any reduction (events \#1 to \#4). The \tell{} of the third one,
\Code{Y \#> Z} gives two value withdrawals, `1' from $D_y$ (\#6) and
`3' from $D_z$ (\#8).  The first reduction modifies the lower bound of
$D_y$ and so wakes the suspended constraint \Code{X \#>= Y}
(\#7). After those two reductions the constraint is suspended and the
waiting one is selected (\#10).  At event \#12, the domains are $D_x =
\{2, 3\}, D_y = \{2, 3\}$ and $D_z = \{1, 2\}$. Then the labelling
phase begins. With our simple ``first fail'' strategy, the first added
constraint is \Code{X \#= 2}.  \Code{X} is ground and equal to 2 and
this constraint is solved (\#17).  Two other constraints are solved
during the propagation, but it leads to $D_y = \{3\}, D_z = \{1, 2\}$
and an empty domain for \Code{X} (\#25).  Another labelling constraint
is tried (\#26), \Code{X \#= 3} and leads to the unique solution
\Code{\{X:3, Y:2, Z:1\}}.

\section{Deriving a Tracer from  the Operational Semantics}
\label{meta:interpreter:section}

In order to experimentally validate the trace defined in
Section~\ref{trace:section}, we derive, from the operational semantics
of Section~\ref{execution:model:section}, a \clpfd{} interpreter in
Prolog that we instrument with trace hooks.  The resulting
interpreter, which produces traces, is not meant to be an efficient
\clpfd{} system, but to be faithful to the semantics of
Section~\ref{execution:model:section}. The faithfulness comes from the
fact that the translation of the semantic rules into executable Prolog
code is syntactical.

In Section~\ref{execution:model:section}, we left the primitive
constraints undefined. We therefore first propose a definition for 8
primitive constraints by specifying, for each, its reduction
operators, its solved condition, and its awakening condition.  Note
that the definitions we propose in
Table~\ref{TAB-CARACTERISTIQUES-CONTRAINTES} define reduction
operators that perform a full-arc consistency.

Then, we show how to translate the primitive constraints and the
semantic rules into Prolog. We also show how to interface this Prolog
code with the Prolog underlying system.

\subsection{Primitive Constraint Definitions}

In order to define a primitive constraint, we need to define its
reduction operators, its solved condition, and its awakening condition
(see Section~\ref{execution:model:section}). We define in
Table~\ref{TAB-CARACTERISTIQUES-CONTRAINTES} the 8 primitive
constraints $x = y$, $x \neq y$, $x = y + n$, $x \neq y + n$, $x > y$,
$x \geq y$, $x = n$, and $x \neq n$, where $x$ and $y$ represent two
finite domain variables, and $n$ represents an integer constant.  The
reduction operator $red_C^x$ is defined by the set of values $W_x$ it
withdraws from the domain of variable $x$.

\begin{table}
  \begin{center}
  \begin{small}
    \begin{tabular}[t]{|l|l|p{3cm}|p{2cm}|}
        \hline
      \multicolumn{1}{|c|}{Constraint} & \multicolumn{1}{c|}
        {Reduction operators} & \multicolumn{1}{c|}{Solved condition} & 
        \multicolumn{1}{c|}{Awakening cond.}\\
      \multicolumn{1}{|c|}{$C$} & \multicolumn{1}{c|}{($red_C^x(\restrict{D}{\Var{c}}) = D_x - W_x$)} 
      & \multicolumn{1}{c|}{$\mathrm{solved\_cond}(C)$} & 
        \multicolumn{1}{c|}{$\mathrm{awake\_cond}(C)$}\\
      \hline \hline

      $x = y$ & $\!\!\begin{array}{l}
        W_x = D_{x} - (D_{x} \cap D_{y})\\
        W_y = D_{y} - (D_{x} \cap D_{y})\end{array}
        $ & $D_{x} = D_{y} = \{v\}$ & $\Cond{x}{any} \vee \Cond{y}{any}$\\
      \hline

      $x \neq y$ & $\!\!\begin{array}{l}
        W_x = \left\{
          \begin{array}{ll}
            \{v\} \cap D_{x}&\mathrm{if\ }D_{y} = \{v\}\\
            \emptyset&\mathrm{otherwise}
          \end{array}\right.\\

        W_y = \left\{
          \begin{array}{ll}
            \{v\} \cap D_y&\mathrm{si\ }D_{x} = \{v\}\\
            \emptyset&\mathrm{otherwise}
          \end{array}\right.
        \end{array}$ & $D_{x} \cap D_{y} = \emptyset$ & $\Cond{x}{ground}~\vee~\Cond{y}{ground}$\\
      \hline

      $x = y + n$ & $\!\!\begin{array}{l}
        W_x = D_{x} - (D_{x} \cap \{v + n, v \in D_{y}\})\\
        W_y = D_{y} - (D_{x} \cap \{v + n, v \in D_{y}\})\end{array}$
      &\begin{minipage}{3cm}{$D_x = \{v_x\}\wedge\\D_y = \{v_y\}\wedge\\v_x = v_y + n$}\end{minipage}
      & $\Cond{x}{any} \vee \Cond{y}{any}$\\
      \hline

      $x \neq y + n$ & $\!\!\begin{array}{l}
        W_x = \left\{
          \begin{array}{ll}
            \{v + n\} \cap D_{x}&\mathrm{if\ }D_{y} = \{v\}\\
            \emptyset&\mathrm{otherwise}
          \end{array}\right.\\

        W_y = \left\{
          \begin{array}{ll}
            \{v - n\} \cap D_{y}&\mathrm{if\ }D_{x} = \{v\}\\
            \emptyset&\mathrm{otherwise}
          \end{array}\right.
        \end{array}$ & $\forall v \in D_{y}, v+n \not\in D_x$ & $\Cond{x}{ground}~\vee~\Cond{y}{ground}$\\
      \hline

      $x > y$ & $\!\!\begin{array}{l}
        W_x = \{v \in D_{x}, v \leq min_{y}\}\\
        W_y = \{v \in D_{y}, v \geq max_{x}\}
      \end{array}$ & $min_{x} > max_{y} $ & $\Cond{x}{max} \vee \Cond{y}{min}$\\
      \hline

      $x \geq y$ & $\!\!\begin{array}{l}
        W_x = \{v \in D_{x}, v < min_{y}\}\\
        W_y = \{v \in D_{y}, v > max_{x}\}\end{array}$ & $min_{x} \geq max_{y} $ & $\Cond{x}{max} \vee \Cond{y}{min}$\\
      \hline

      $x = n$ & $\!\!\begin{array}{l}
      W_x = D_{x} - \{n\}\end{array}$ &
      $D_{x} = \{n\}$ &
      \\
      \hline

      $x \neq n$ & $\!\!\begin{array}{l}
      W_x = \left\{\begin{array}{ll}\{n\}&\mathrm{if\ }n\in D_{x}\\\emptyset&\mathrm{otherwise}\end{array}\right.\end{array}$ &
      $n \notin D_{x}$ &
      \\
      \hline

    \end{tabular}
    \caption{Characteristics of the  primitive constraints implemented
    in the interpreter}
    \label{TAB-CARACTERISTIQUES-CONTRAINTES}
  \end{small}
  \end{center}
\end{table}

The equality constraint between two variables $x$ and $y$ ($x = y$)
withdraws from the domains of $D_x$ and $D_y$ the values which are not
contained in both domains.  The constraint is solved only when the two
variables are ground and have the same value $v$. At each modification
of $D_x$ or $D_y$, the reduction operators may withdraw new
values. The constraints must therefore be woken at each of their
modification.

The difference constraint between two variables $x$ and $y$ ($x \neq
y$) can only reduce the domains when one of them is ground. However,
it is solved as soon as the two domains are disjoint. 

\subsection{Data Structures}

A constraint variable is represented by a term containing a unique
integer, and a string (its name in the source).
A constraint instance is represented by a quadruple containing a unique
constraint number, a string (the constraint as displayed in the
source, i.e., its abstract representation), an internal (concrete) form, a list of constraint variables, and an
invocation context. The invocation context of a constraint is the
Prolog goal from which it was invoked.
The solver state is represented by a sextuplet: \Code{(A, S, Q, T, R,
  D)}, where \Code{A}, \Code{S}, \Code{Q}, \Code{T} and \Code{R}
define the store as described in Section~\ref{execution:model:section};
they are lists of constraints. \Code{D} is a list of domains.

In order to represent awakening conditions and domain narrowing, we
define five types of domain modifications (following what is done in
Eclipse~\cite{eclipse}).  Each type refers to a particular constraint
variable $x$.
\begin{myitemize}
\item $x_{min}$ refers to a modification of the $D_x$  lower bound,
e.g., $\{1, 2, 4\} \rightarrow \{2, 4\}$;
\item $x_{max}$ refers to a modification of the $D_x$  upper bound,
e.g., $\{1, 2, 4\} \rightarrow \{1, 2\}$;
\item $x_{any}$ refers to any modification of $D_x$,
 e.g., $\{1, 2, 4\}~\rightarrow~\{1, 4\}$;
\item $x_{ground}$ refers to a grounding of $x$ ($D_x$ becomes a
singleton), e.g., $\{1, 2, 4\}~\rightarrow~\{1\}$;
\item $x_{empty}$ refers to an emptying of $D_x$, e.g., $\{1, 2,
  4\}~\rightarrow~\emptyset$
\end{myitemize}

Those five modification types respectively appear in the code as
\Code{X->min}, \Code{X->max}, \Code{X->any}, \Code{X->ground}, and
\Code{X->empty}.  Awakening conditions are disjunctions of such
modification types; such disjunctions are encoded by lists.

\subsection{Translation of the Primitive Constraints}

The reduction operators, the solved condition, and the awakening
condition defining a primitive constraint are encoded by the
following predicates:

\begin{myitemize}
\item \Code{cd\_reduction(+C, +D, +X, -Wx)}\footnote{As specified
in the standard Prolog~\cite{alipie96}, {\tt +} denotes input arguments and
{\tt -} denotes outputs arguments.}: takes as input a
  constraint $C$, a domain state $D$, and a constraint variable $x$; it
  succeeds iff the application of $red^{x}_{C}(\restrict{D}{\Var{C}})$
  withdraws a non-empty set (bound to \Code{Wx}). There is one
  clause per reduction operator of \Code{C};
  
\item \Code{cd\_solved(+C, +D)}: takes as input a constraint $C$ and
  a domain state $D$; it succeeds iff the constraint $C$ is solved in
  the domain state $D$;
  
\item \Code{cd\_awake(+C, -Cond)}: takes as input a constraint $C$
  and outputs (in \Code{Cond}) the list of awakening conditions of
  $C$.
\end{myitemize}
 
\begin{figure}[t]
\begin{minipage}[t]{0.45\linewidth}\small
\begin{listing}[1]{1}
cd_reduction(diff(X,Y),D,Y,[Vx]) :-
  is_ground(X,D,Vx),
  get_domain(Y,D,Dy),
  member(Vx, Dy).
cd_reduction(diff(X,Y),D,X,[Vy]) :-
  is_ground(Y,D,Vy),
  get_domain(X,D,Dx),
  member(Vy,Dx).
\end{listing}
\end{minipage}\hfill
\begin{minipage}[t]{0.45\linewidth}\small
\begin{listing}[1]{9}
cd_awake(diff(X,Y),Cond) :-
  Cond = [X->ground,Y->ground]).

cd_solved(diff(X,Y),D) :-
  get_domain(X,D,Dx),
  get_domain(Y,D,Dy),
  d_intersection(Dx,Dy,[]).
\end{listing}
\end{minipage}
\begin{center}
\caption{Definition  of the primitive constraint
  $ x\neq y$: its two reduction operators, its solved condition, and
  its awakening condition, as specified in
  Table~\ref{TAB-CARACTERISTIQUES-CONTRAINTES}.}
\label{PRG-DIFF}
\end{center}
\end{figure}

Figure~\ref{PRG-DIFF} shows the implementation of the primitive
constraint $ x\neq y$ which is simply a Prolog encoding of the first
entry of Table~\ref{TAB-CARACTERISTIQUES-CONTRAINTES}.
The functor \Code{diff/2} is the internal encoding of $\neq$.
Predicate \Code{is\_ground(+X, +D, -Vx)} takes a constraint variable $x$ and a
domain $D$, and succeeds iff $D_x$ is a singleton (bound to
\Code{Vx}); \Code{get\_domain(+X, +D, -Dx)} takes a
constraint variable $x$ and a domain state $D$, and outputs the domain of
$x$ ($D_x$); \Code{d\_intersection(+D1, +D2, -D)} computes the
intersection of two domains.

\subsection{Translation of the Semantic Rules}

Figure~\ref{meta:interpreter:figure} contains the translation of the
semantic rules of Figures~\ref{FIG-RULES} and~\ref{FIG-RULES2}. 
Each rule is encoded by a predicate with the same name as the rule.
The translation is merely syntactical, except for the \tell{} and
\told{} rules, for which it is unnecessary to save and restore the
solver states ($push$ and $pop$) since this work is done by the
Prolog backtracking mechanism.

Before paraphrasing the code for one rule, we give the meaning of all
the (simple) predicates that are not defined elsewhere in the paper:
\Code{choose\_in\_queue(+Q0, -C)} takes as input a queue $Q$ and
outputs one of the queue constraints ; it succeeds iff $Q$ is not
empty.  The choice of the constraint depends of the solver strategy;
\Code{subtract(+L1, +L2, -L)} computes the difference between two
lists;
\Code{get\_varC(+C, -V)} takes as input a constraint $C$ and outputs
a list of the constraint variables that appear in $C$;
\Code{trace(+Port, +C, +St0, +O1, +O2, +O3)} takes as input the
different event attributes as described in
Section~\ref{trace:section}; it calls the trace analysis system which
can, for example, print a trace line;
\Code{put\_end\_of\_queue(+C, +Q0, -Q)} puts a constraint at the end
of a queue;
\Code{update\_domain(+X, +Wx, +D0, -D, -Mod)} takes as input a
constraint variable $x$, a value set $W_x$ (to withdraw), and a
domain state $D^0$; it outputs the state domain $D$ such that
$D_x=D^0_x - W_x$, and the list of modification types $x_{mod_1}$,
..., $x_{mod_n}$ (where $mod_i \in \{min, max, ground, any, empty\}$)
that characterizes  the $W_x$ value removals;
\Code{internal(+C, -Ci)} takes as input a constraint and outputs its
internal representation.

All the predicates translating rules take as input a solver state
$St_0$ and output a new solver state $St$.  $St_0$ and $St$
respectively denote the state of the solver before and after the
application of a rule.
The only exceptions are predicates \Code{wake\_up(+St0, -St, +ModIn)}
and \Code{reduce(+St0, -St, -ModOut)} that respectively inputs and
outputs an additional argument: a list of modification types
($x_{mod_1}$, ..., $x_{mod_n}$).  This list is computed in
\Code{reduce/3} and used in \Code{wake\_up/3} to check the awakening
condition.

Predicate \Code{select(+St0, -St)} translates the {\bf select} rule.
That rule needs to fulfill 3 conditions to be allowed to be applied:
$\exists C \in Q$, that is checked line~17 by
\Code{choose\_in\_queue/2} (which fails iff the queue is empty); $A =
\emptyset$ and $R = \emptyset$ that are checked line 15. The 2 actions
to perform when the conditions of the rule hold are $Q \gets Q -
\{C\}$, which is done line 18 by \Code{subtract/3}, and $A = \{C\}$
which is done line 16.
The other rules are translated in the same way.

\begin{figure}
\begin{minipage}[t]{.45\linewidth}\small
\begin{listing}[1]{1}
tell(C, St0, St) :-
  St0 = ( [], S, Q, T, R, D),
  St  = ([C], S, Q, T, R, D),
  trace(tell, C, St0, -, -, -).

told(C, St0) :-
  trace(told, C, St0, -, -, -).

select(St0, St) :-
  St0 = ( [], S, Q0, T, [], D),
  St  = ([C], S,  Q, T, [], D),
  choose_in_queue(Q0, C),
  subtract(Q0, [C], Q),
  trace(select, C, St0, -, -, -).

reject(St0, St) :-
  St0 = ([C], S, Q, T,  [], D),
  St  = ( [], S, Q, T, [C], D),
  get_varC(C, VarC),
  member(X, VarC),
  get_domain(X, D, []),
  trace(reject, C, St0, -, -, -).

wake_up(St0, St, ModIn) :-
  St0 = (A, S0, Q0, T, [], D),
  St  = (A, S,   Q, T, [], D),
  member(C, S0),
  awake_cond(C, ModIn, True),
  subtract(S0, [C], S),
  put_end_of_queue(C, Q0, Q),
  trace(wake_up, C, St0, True, -, -).
\end{listing}
\end{minipage}\hfill
\begin{minipage}[t]{.45\linewidth}\small
\begin{listing}[1]{37}
reduce(St0, St, ModOut) :-
  St0 = ([C], S, Q, T, [], D0),
  St  = ([C], S, Q, T, [], D),
  get_varC(C, VarC),
  member(X, VarC),
  reduction(C, D0, X, Wx),
  update_domain(X, Wx, D0, D, ModOut),
  trace(reduce, C, St0, X, Wx, ModOut).
 
true(St0, St) :-
  St0 = ([C], S, Q, T0, [], D),
  St  = ( [], S, Q, T,  [], D),
  solved_cond(C, D),
  T = [C|T0],
  trace(true, C, St0, -, -, -).

suspend(St0, St) :-
  St0 = ([C], S0, Q, T, [], D),
  St  = ( [], S,  Q, T, [], D),
  S = [C|S0],
  trace(suspend, C, St0, -, -, -).

reduction(C, D, X, Wx) :-
  internal(C, Ci),
  cd_reduction(Ci, D, X, Wx).

awake_cond(C, ModIn, True) :-
  internal(C, Ci),
  cd_awake(Ci, Cond),
  intersection(Cond, ModIn, True),
  not True = [].

solved_cond(C, D) :-
  internal(C, Ci),
  cd_solved(Ci, D).
\end{listing}
\end{minipage}
\begin{center}
  \caption{Prolog translation of the semantic rules of Figures~\ref{FIG-RULES} and~\ref{FIG-RULES2}}
  \label{meta:interpreter:figure}
\end{center}
\end{figure}

\begin{figure}
\begin{minipage}[t]{.45\linewidth}\small
\begin{listing}[1]{1}
call_constraint(C, St0, St) :-
  tell(C, St0, St1),
  propagation(none, St1, St),
  ( true ; told(C, St), fail ),
  St = (_, _, _, _, [], _). 

propagation(Mod0, St0, St) :-
  prop_step(St0, St1, Mod0, Mod)
  -> ( St1 = (_, _, _, _, [], _)
       -> propagation(Mod, St1, St)
       ;  St = St1 )
  ;  St = St0.
\end{listing}
\end{minipage}\hfill
\begin{minipage}[t]{.45\linewidth}\small
\begin{listing}[1]{13}
prop_step(St0, St, Mod0, Mod) :-
    select(St0, St)  -> Mod = none
;   reject(St0, St)  -> Mod = none
;   wake_up(St0, St, Mod0) 
                     -> Mod = Mod0
;   reduce(St0, St, Mod) 
                     -> true
;   true(St0, St)    -> Mod = none
;   suspend(St0, St) -> Mod = none.
\end{listing}
\end{minipage}
\begin{center}
  \caption{Integrating the  constraint solver with the underlying Prolog system}
  \label{meta:interpreter:control:figure}
\end{center}
\end{figure}

\subsection{Integration with the Underlying Prolog  System}

The integration of our instrumented constraint solver with the
underlying Prolog system is done by the predicate
\Code{call\_constraint/3} which is given in
Figure~\ref{meta:interpreter:control:figure}.  After propagation and
success, it returns the new state of the constraint part (\Code{St}).
If the propagation leads to a failure, the goal fails.

Predicate \Code{prop\_step/4} performs a propagation step, i.e., it applies one
of the 6 propagation rules of Figure~\ref{FIG-RULES}; it fails if
no rule can be applied.  The choice of the rule to apply is done
according to the strategy discussed in
Section~\ref{propagation:section}.
Predicate \Code{propagation/3} calls \Code{prop\_step/4} in loop
until either a propagation step fails (the fix-point is reached) or
the solver rejects a constraint (the constraint goal is
unsatisfiable).

\section{Experimentation}
\label{validation:section}

\begin{figure}
\begin{center}
\includegraphics*[scale=0.5]{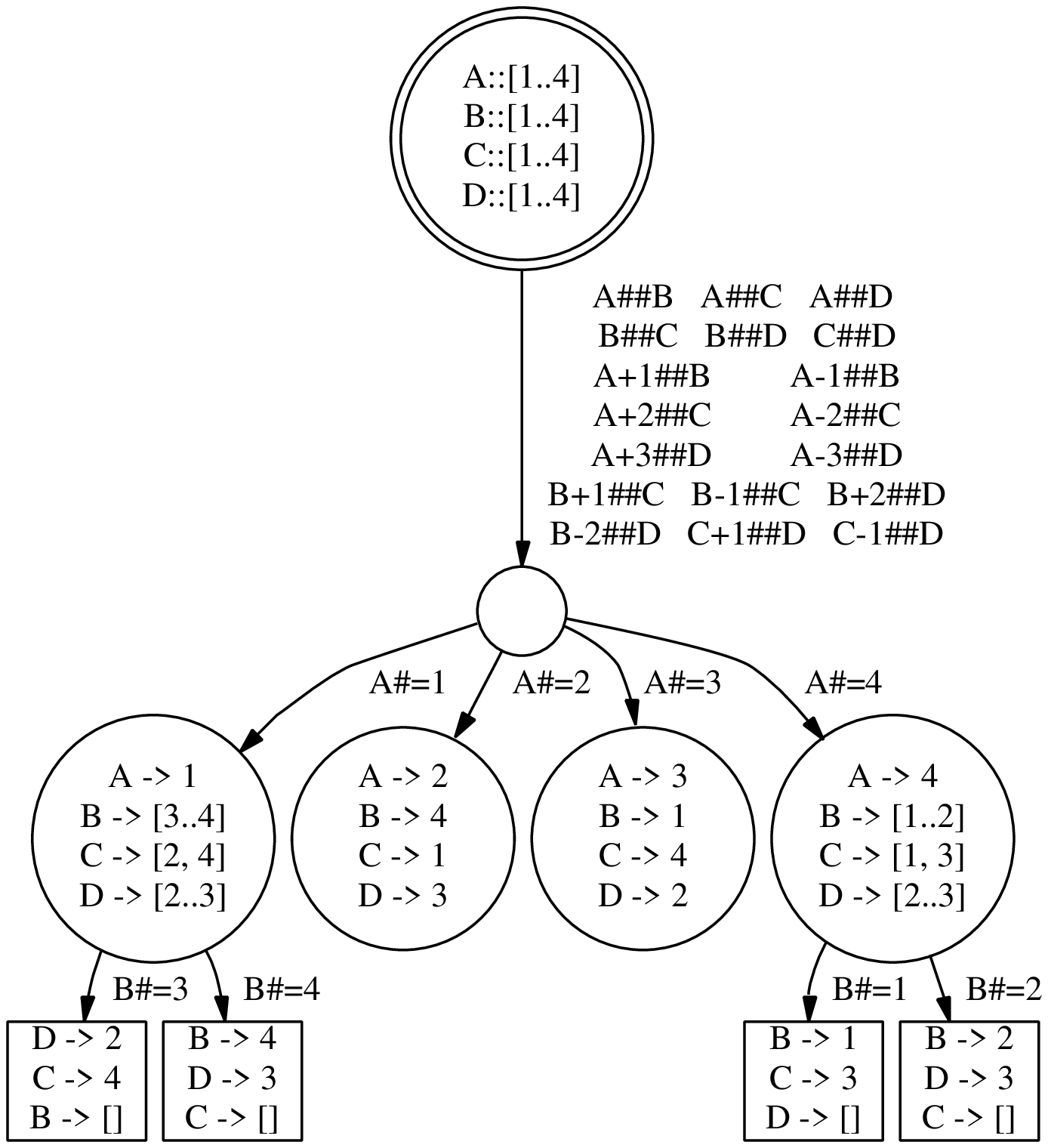}
\end{center}
\caption{A search-tree of a 4-queens constraint program execution,
obtained by trace analysis, and displayed by \emph{dot}.}
\label{search:tree:fig}
\end{figure}

We have made some preliminary experiments of trace analysis with our
tracer.

A first experiment consists in displaying the labelling (or search-)
tree.  The analyzed program solves the well known n-queens problem.
Figure~\ref{search:tree:fig} shows a search-tree obtained for the
4-queens problem by trace analysis of a trace generated by our
meta-interpreter, and displayed with \emph{dot}~\cite{koutsofios91}.
Each variable corresponds to a row of the chess-board.  The top node
of the displayed tree contains the initial domains, the other nodes
contain the reduced domains if any.  The tree shows four failures in
square boxes, two solutions and three choice points. The uppest arc
sums up the \tell{} operations for all permanent constraints of the
programs. Each other arc represents a labelling constraint addition.

The trace analysis uses only the \tell{}, \told{}, \reduce{} events.
Few attributes were needed: the port, the concerned constraint and the
domains of constraint variables. For \reduce{} events, the updated
variable and the withdrawn set were also needed.

The sequence of \tell{} and \told{} events in the trace corresponds to
a depth-first left-to-right visit of the search-tree where \tell{} and
\told{} events respectively correspond to downward and upward
moves in the search-tree. Reconstructing a tree from its visit (for a
fixed visit strategy) is easy.  In Figure~\ref{search:tree:fig},
we took advantage of \reduce{} events to label the nodes with the
propagation results.

\begin{figure}
\begin{center}
        \includegraphics[width=6.5cm]{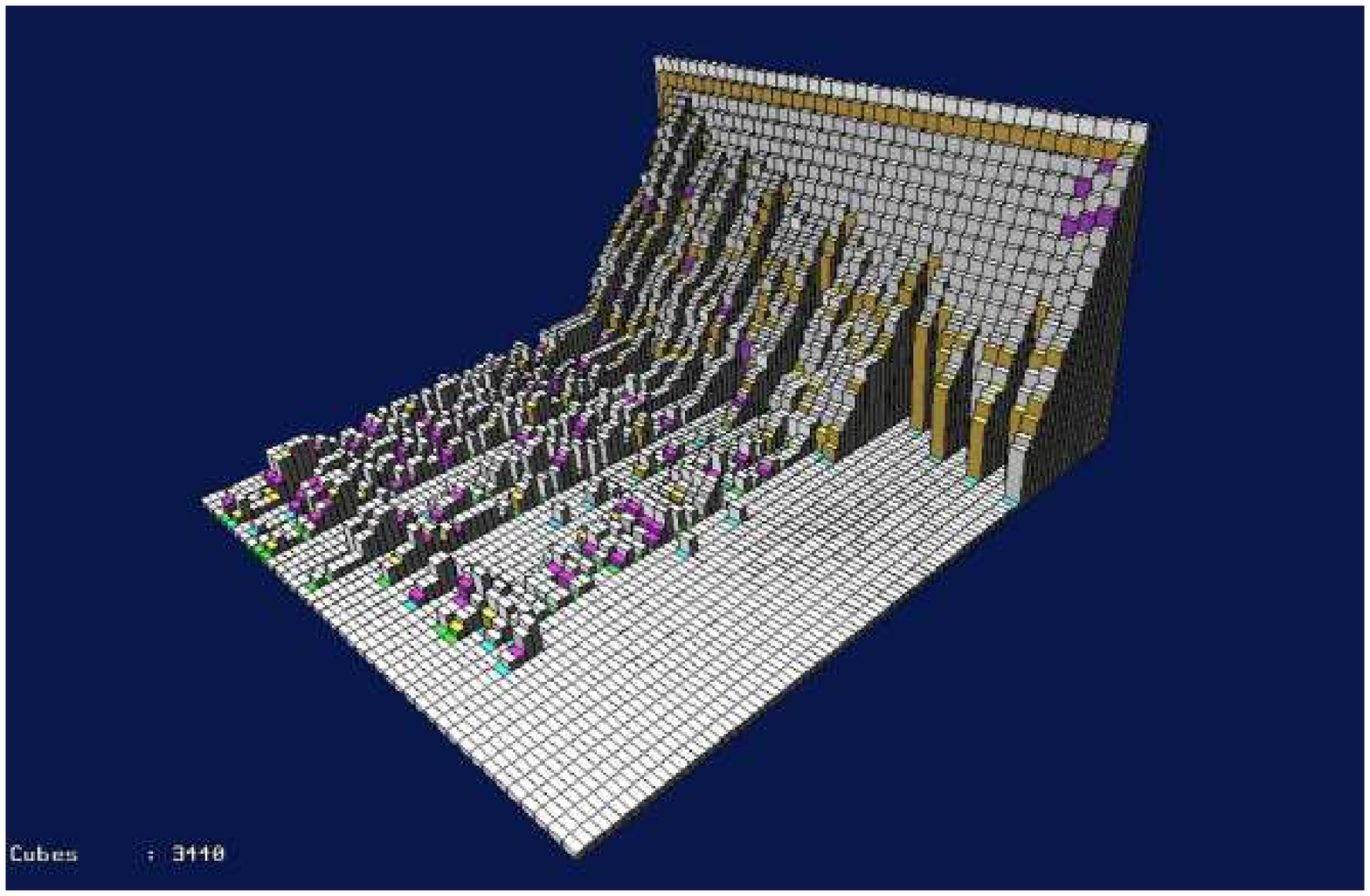}
        \hfill
        \includegraphics[width=6.5cm]{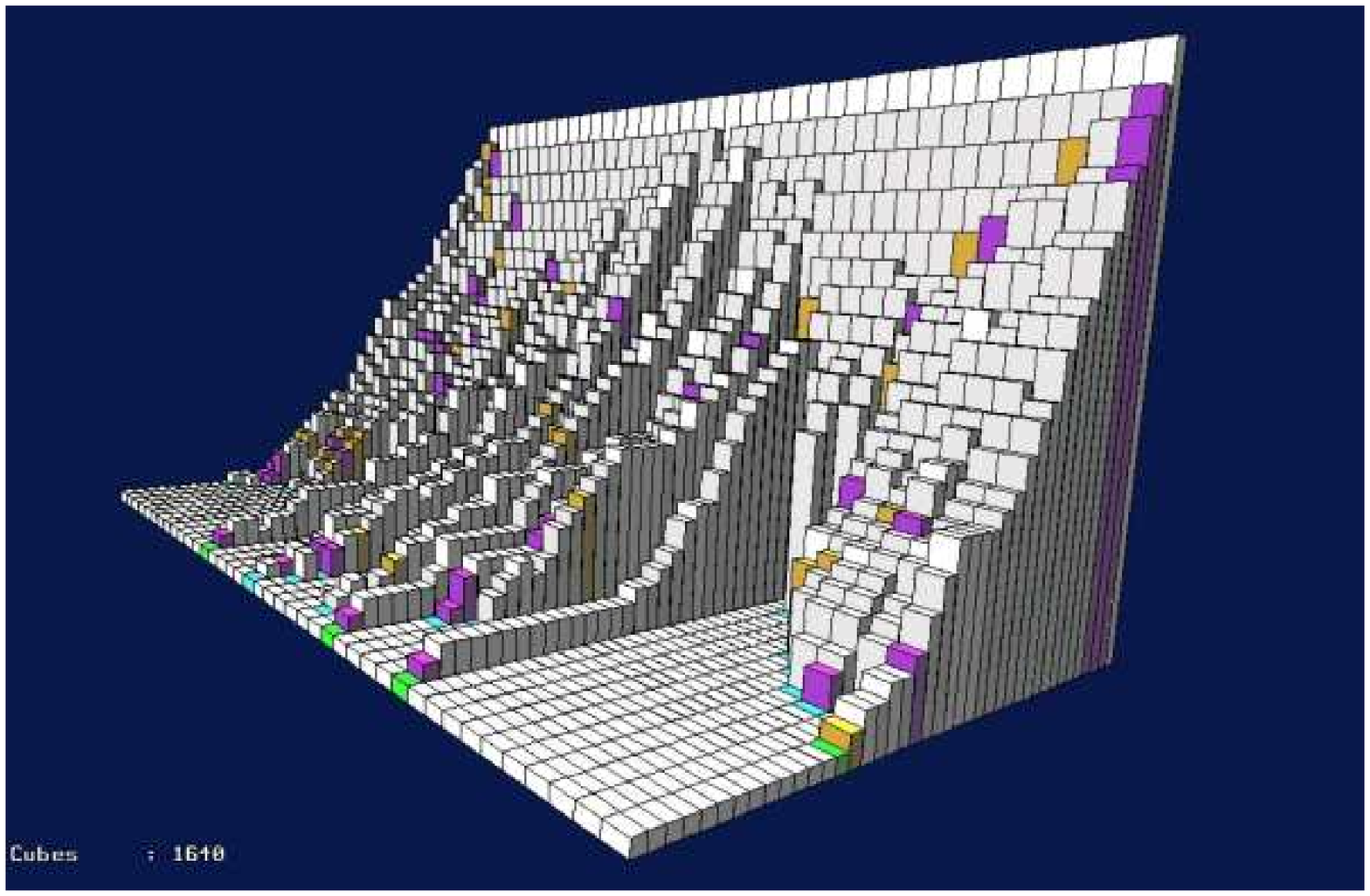}
\caption{Comparing two search procedures for the 40-queens problem with VRML
views computed by trace analysis.
}\label{FIG-VRML}
\end{center}
\end{figure}

\paragraph{}
A second experiment is the generation of a 3D variable update view
\cite{zoumman01}.  The evolution of the domains of the variables
during the computation is displayed in three dimensions. It gives a
tool \emph{à la} TRIFID~\cite{mancarro00} (here however colors are
introduced to display specific events as in the variable update view
of~\cite{simonis00}).  The trace analyzer makes a VRML file by
computing domain size on each \tell{} and \reject{} event, and when a
solution is found. The details of \reduce{} events allow us to assign
color to each kind of domain update (for example minimum or maximum
value removed or domain emptied) as made by Simonis and Aggoun in the
Cosytec Search-Tree Visualizer~\cite{simonis00}.  The trace analysis is
implemented in about 125 lines of Prolog and generates an intermediate
file. A program implemented in 240 lines of C converts this file into
VRML format.

Figure~\ref{FIG-VRML} shows the resolution of the 40-queens problem
with two different labelling strategies. We have three axes: variables
(horizontal axis of the vertical ``wall''), domain size (vertical
axis) and time. The first strategy is a first-fail selection of the
labelled variable and the first value tried is the minimum of its
domain. The second strategy is also a first-fail strategy but variable
list is sorted with the middle variable first and the middle of domain
is preferred to its minimum.  This strategy derives from one described
by Simonis and Aggoun~\cite{simonis00}.  This approach allows to
compare the efficiency of these two strategies by manipulating the
3D-model. With the first strategy, we get a quick decreasing of the
domain size on one side of the chess-board and a long oscillation of
the domain size on the other side. With the second strategy, the
decreasing of domain size is more regular and more symmetrical, the
solution is found faster. In fact, the second strategy, which consists
in putting the queens from the center of the chess-board, benefits
more from the symmetrical nature of the problem. The possibility of
moving manually the figure facilitates observation of such property.

\section{Discussion and Conclusion}
\label{conclusion:section}

This article is a first attempt to define new ports for tracing finite
domain solvers. We do not claim that the trace model presented here is
the ultimate one. On the contrary we propose a methodology to
experiment and to improve it. This methodology is based on the
following steps: definition of an executable formal model of trace,
extraction of relevant informations by a trace analyzer, utilization
of the extracted informations in several debugging tools. Each
debugging tool extracts from the same generic trace the information
it needs.

The formal approach of trace modeling used here allows to clearly
define the ports. Then the implementation of the model by a
meta-interpreter almost written in ISO-Prolog, a logic programming
language with a clear semantics, allows to preserve its correctness
wrt the formal model.  This methodology is efficient enough on small
examples and therefore is of practical interest. However to handle
large realistic examples will require hard-coded implementation of the
trace model.

Different aspects must be examined to estimate the results. First the
formal model itself. Two ports are directly related to logic
programming (\tell{} and \told{}) and correspond to the well-known
ports {\em call} and {\em exit} of \cite{byrd80}, the others
correspond to a small number of different steps of computation of the
reduction operator fix-point. On the other side we defined a great
number of (possibly large sized) attributes to ensure that each event
carries enough potentially useful informations.  This immediately
shows the limits of our model: it is probably general enough to take
into account several finite domain solvers, but tracing the complete
behavior of different solvers will require new or different ports to
take into account different kinds of control, specific steps of
computation (e.g. constraint posting, labelling phase, \ldots), or
different algorithms.

The proposed methodology
shows the way to progress: defining the trace with a formal model
makes easier to compare different trace models. It is thus
easier to see which are the missing ports or attributes.
Some solvers use a propagation queue with events instead of
constraints (in this case new ports or same ports  but with
different attributes are necessary), or do not use backtracking (then
\told{} is never used).
The question is also to find the right balance between the number of
events and the attributes, in such a way that a hard-coded efficient
implementation of (a part of) the trace model is still possible.
For example the attribute {\tt withdrawn} of the port \reduce{}
concerns one variable only. Therefore if several variables are
involved by a single reduction step, there will be several
\reduce{} events. Another possibility would be to have a unique
event with a port whose {\tt withdrawn} attribute includes several
variables at once. Gathering too many informations in one event may
slow down the tracer considerably. The trace production must be as
fast as possible in order to keep the best performances of the
solver.

The ports and attributes presented here seem to be a good basis to
start the study of a generic trace for \clpfd{}. In \cite{oad2221} a
sample of additional ports are suggested to cope with different finite
domain solvers.

The second aspect concerns the extraction of tool specific relevant informations. 
The generic trace, by definition, contains too many informations if
used by a unique debugging tool only. It is also not intended to be
stored in a huge file, but it should be filtered on the fly and re-formatted for use
by some given tool. The methodology we proposed here,
\emph{à la opium}, is well-known and has been shown to be efficient and
general enough to be used in practice also in hard-coded
implementations. It allows to specify the trace analysis in a high
level language (here Prolog) in a way which is independent from the trace
production.

For each experimented tool we have presented here
there is a specific analyzer which needs only few lines of code. With
this approach, building different views of the same execution requires
only to modify the trace analyzer. Notice that the implementation 
presented here assumes that there is only one trace analyzer running
in parallel with the solver. The question of how different
analyzers, working simultaneously, could be combined is not considered here.

Finally the third aspect concerns the experimentation. The main challenge in
constraint debugging is performance debugging. Our objective is to
facilitate the development of constraint resolution analysis tools in
a manner which is as independent as possible from the solvers
platforms. The three steps method (generic tracer/trace
analyzer/debugging tool) is a way to approach such a goal. 
We experimented it by building several analyzers and (limited) tools
very easily, without having to change the trace format. Of course much
more experiences are still needed.

Another way to estimate the results of this article is to consider some of
the existing debugging tools and to observe that most of the
informations relevant for each of these tools are already present in the ports
and their attributes. Here are briefly reviewed some platform
independent tools\footnote{They are considered as (\clpfd{}) platform
  independent tools in the sense that they are focused on general
  properties of finite domain solving: choice-tree, labelling,
  variables domain evolution.}, namely debugging tools using graphical interfaces 
where labelling, constraints and propagation can be visualized.

The Search-Tree Visualization tool for CHIP, described by Simonis and
Aggoun~\cite{simonis00} displays search-trees, variables and domain
evolution. The whole of this information is present in the proposed
ports.  Our experimentation shows that search-tree and constraints
can easily be displayed.  The relevant information is also present for
the Oz Explorer, a visual programming tool described by
Schulte~\cite{schulte97}, also centered on the search-tree
visualization with user-defined displays for nodes.  It is also the
case in the search-tree abstractor described by Aillaud and
Deransart~\cite{aildera00} where search-tree are displayed with
constraints at the nodes.

In Grace, a constraint program debugger designed by
Meier~\cite{meier95}, users can get information on domain updates and
constraint awakenings. This is available in our trace via our 8
ports. Grace also provides the ability to evaluate expressions using
the current domain of the variables. As the domain of the variables is
an attribute of all events, it can be obtained by any analyzer of
our trace.  The CHIP tool also provide update views in which, for
example, useless awakenings are visible. In our environment useless
awakenings could be detected by a \select{} not followed by a
\reduce{}. The visualization tool by Carro and
Hermenegildo~\cite{mancarro00} traces the constraint propagation in 3
dimensions according to time, variable, and cardinal of the domain. The
required information is present in our trace.  The S-boxes of Goualard
and Benhamou~\cite{goualard00} structure the propagation in and the
display of the constraint store.  Structuring the propagation requires
to modify the control in the solver.  Displaying the store according
to the clausal structure would however cause no problem.

An important aspect, not presented in this paper, concerns the
possibility of interactive tools. This means that some debugging tools
may request to re-execute some already traced execution. Special
mechanisms must be involved in order to re-execute efficiently and
with the same semantics some part of the execution. How such
capability may influences the form of the generic trace is still an
open question and will be further investigated.


\end{document}